# Service oriented infrastructure (Part 2)

## Securing business operations in an SOA

T Dimitrakos and D Brossard (BT) and P de Leusse (Newcastle University)

*Service-oriented infrastructures pose new challenges in a number of areas, notably with regard to security and dependability.*

*BT has developed a combination of innovative security solutions and governance frameworks that can address these challenges. They include advances in identity federation; distributed usage and access management; context-aware secure messaging, routing and transformation; and (security) policy governance for service-oriented architectures.*

*This paper discusses these developments and the steps being taken to validate their functionality and performance.*

IT and communication service providers and their corporate customers are increasingly introducing Service-Oriented Architectures (SOAs) to cut costs, enhance their agility and reduce time-to-market. Service-Oriented Infrastructures (SOIs) amplify such benefits. In contrast to traditional infrastructures, in which resources that were scaled to meet peak demand were dedicated to particular applications on a permanent basis, SOIs exploit virtualisation to the full, allocating resources to applications in a way that constantly matches supply with demand.

Simultaneously, the ways in which organisations manage their affairs are changing. Their workforces are much more mobile, for example, and suppliers and outsourcing partners play much bigger roles in the delivery of their products and services. Increasingly, organisations also want customers to be able to serve themselves, interacting directly with corporate IT systems to make purchases, report faults and so on.

To support such new ways of doing business, organisations must make their IT systems available beyond their corporate networks. It isn't just a matter of allowing customers, suppliers and partners to log on and use whichever IT systems they need to complete their tasks. Increasingly, those who work together will also want to integrate their infrastructures and applications so that policies and data relevant to their relationships can flow securely between them in accordance with their agreements. Together, these developments pose new challenges in the areas of security and compliance.

On the one hand, the incidence of attacks that exploit networked computing power and collaboration technology to gain access to corporate IT systems, steal information and cause damage has been increasing. It is reasonable to expect the problem to grow further as the use of distributed SOIs becomes commonplace. On the other hand, factors such as distributed ownership can make attacks particularly difficult to detect and address: as noted in [1], malicious intent is often only recognisable as an emerging property of the network. This is an issue that clearly needs to be addressed.

Once threats have been identified, an immediate and coordinated response is essential. Changes may need to be made to both usage and access policies and business process parameters to mitigate the risks involved. Cross- and intra-enterprise compliance will be equally critical when it comes to ensuring compliance. Legal and regulatory frameworks are becoming more complex and less forgiving. Organisations must therefore comply not only with the laws and regulations that apply where they operate, but also with those that apply to their clients and partners.

The involvement of multiple organisations creates complex relationships, especially with regard to the ownership of resources and information. Each of the organisations using an SOI will want to define its own policies governing entitlements, the usage of resources and access to them, for example. They will also want to know how these policies have performed at any given time – past, present or future. Unfortunately, however, their visibility of the processes in which they participate and their consequences may be limited. As a result, it becomes much harder for organisations to govern their relationships with customers, suppliers and partners in ways that are safe and controlled. It won't always be clear how their information and resources are used across the value chain and: that may make it difficult for them to identify and assess the impact of violations of their policies or agreements.

The emergence of virtual organisations complicates the situation further. Such coalitions of individuals, groups, organisational units or even entire businesses may only exist for short periods, pooling their resources, capabilities and/or knowledge to pursue some shared objective. If they need new infrastructure, it must be put in place quickly. With little time available to negotiate details such as security and compliance procedures, those involved will be looking for 'out of the box' solutions that are scalable, responsive and adaptable.

To address these problems, organisations will depend on interdisciplinary approaches that draw as much on expertise in law, economics and business management as on more obvious disciplines such as telecommunications and grid or 'cloud' computing.

Over the past five years, BT's researchers have been working with academics and industrial partners to develop the solutions required[1]. This paper reviews the approaches they have developed and describes how they are being applied in the context of BT's SOI research programme.

1.1 Overview of security capabilities

The essence of an SOI is the delivery of ICT infrastructure (i.e., compute, storage and network) as a set of services. We take as our starting point the three-layer model presented in figure 1, which is introduced in [2] and explained in detail in [3]. Ways in which organisations can use SOIs are described in [2]. Figure 2 illustrates a suite of capabilities serviceoriented enterprises can use to secure their networked IT infrastructures. Those highlighted (in dark gray) are discussed in this paper.

In section 2, we discuss the secure messaging and application gateways in figure 2. Section 3 then discusses federated identity management and identity brokerage, while section 4 discusses access management. In the case of SOIs operated by service providers like BT, the protocols customers must use and the conditions under which interactions with services can occur are made available to users through declarative policies and agreements. In the bottom layer of figure 2, a security enforcement layer is shown. This layer consists of a network of security enforcement points distributed across the SOI. These can be embedded in service gateways, message brokers and web application servers. They implement the message interceptor, message inspector, message broker and service proxy design patterns to allow the enforcement of actions for service endpoints independently of the application logic. Enforcement is based on sets of rules that can be specified as declarative policies that are private to a service exposure. These policies specify behaviour that focuses on non-functional requirements and therefore complements the business application logic, which focuses on meeting the service's functional requirements.

The performance and monitoring of enforcement actions often depends on the support of separate infrastructure services, such as the ones shown in the middle layer of figure 2. The choice of the actions to be enforced, and of the way these are enforced, may depend on the content of the messages exchanged, their correlation pattern, the context of a transaction, the security claims (for example, with respect to identity, attributes and credentials) of the requestor and the authorisation policies in place. The middle layer of figure 2 shows value-adding 'identity brokerage' services that empower the enterprise to define how identity and personal information is disclosed and managed in different contexts, such as in different value chains and different business-to-business collaborations. An advanced, next-generation identity-brokerage service for business-tobusiness collaborations is presented in section 3 of this paper.

The middle layer of figure 2 also shows value-adding services for 'usage and access control' that enable enterprises to manage access to their resources and the entitlements of user communities in different collaboration contexts. In section 4, we present an advanced authorisation service that is capable of managing usage and access in multi-administrative environments, such as those that appear in multi-tenancy hosting and in business or government coalitions.

The middle layer of figure 2 also includes a security dashboard that aggregates services focusing on security analytics and an autonomics layer. These services correlate events (including reports of violations of security policy), analyse dependences, identify and report possible risks and (possibly) recommend measures to mitigate them, including possible reconfigurations of the security infrastructure or changes of security policy.

The top layer of figure 2 refers to a governance layer that manages (a) the life-cycle of a secure exposure of business services, (b) the composition of such services with a collection of security capabilities that implement nonfunctional requirements and (c) the life-cycle of policies associated with each security capability.

The detail of any changes to policy enforcement may depend on the nature of the transaction, the agreements that are in place and events occurring elsewhere in the infrastructure. In addition, both the enforcement logic and the 'semantics' of enforcement actions may have to be updated. In the latter case, changes could affect both the operation of the encapsulated components (that is, those performing an enforcement action) and external infrastructure service dependencies (that is, those associated with the enforcement of an action).

Whatever updates are required will need to be coordinated across the infrastructure. Consistency between the 'private' enforcement logic, 'public' policies and the agreements between service providers and users must be maintained.

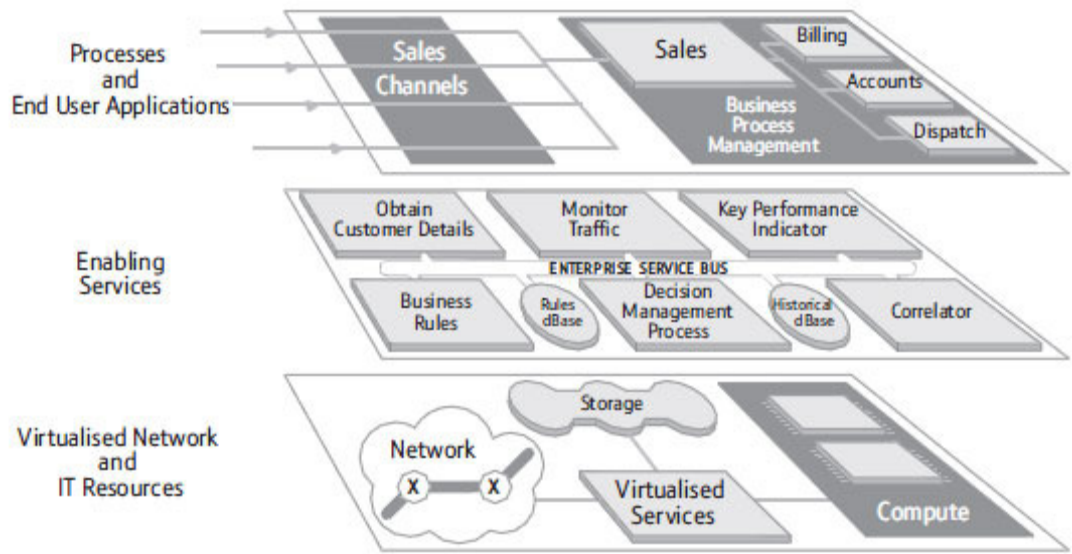

Figure 1. The three level model for a service-oriented infrastructure

Web services, which play a fundamental role in SOAs, are often based on the use of the eXtensible Markup Language (XML). This makes it possible for applications to interact over almost any transport protocol, including common web protocols such as HTTP. However, it also makes it possible for messages to carry harmful content past traditional security guards. For example, messages could be malformed in ways that cause parsers and applications to malfunction. However, many traditional network firewalls lack the ability to inspect XML messages, validate their structures, check them against service Application Programming Interfaces (APIs) and detect anomalous or suspicious content. Similarly, in typical SOA deployments, messages must pass through a multitude of intermediaries, each of which may require some visibility of the message and may be expected to perform some message processing actions. This challenges network security technologies such as Secure Sockets Layer (SSL), Internet Protocol Security (IPsec) and Virtual Private Networks (VPNs) that were designed to ensure point-to-point security and, as a result, aren't able to preserve the integrity and privacy of content as messages pass between message processing intermediaries and other applications on their way from one organisation to another. Neither can they provide messagelevel audit trails or ensure end-to-end non-repudiation.

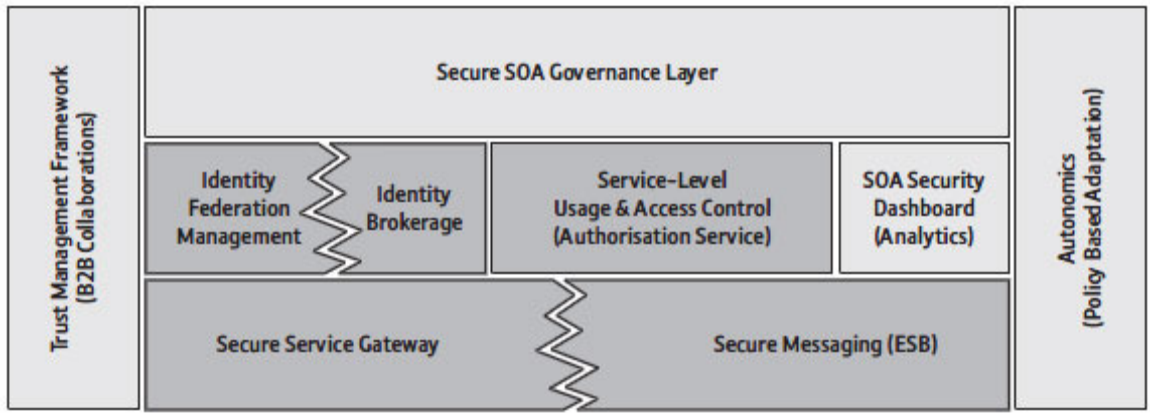

Figure 2. Overview of the security capabilities required by service-oriented enterprises

A common way of addressing such problems has been to program XML and web services security directly into application-based services. However, this is dependent on the availability of highly-skilled developers who understand emerging XML and WS-* web service security standards and know how to implement them effectively. Another problem is that security policies have to be implemented repeatedly on different platforms and maintained throughout the lifetimes of the applications involved. As well as driving up the cost of implementing and managing the enforcement security policy, this increases the probability that vulnerabilities will be caused by implementation errors and platform limitations. Such vulnerabilities can be very difficult to detect and fix once they have been introduced. In addition, the developers in the various organisations involved must somehow coordinate security policies and implementations. Web services cannot communicate security expectations or capabilities to clients automatically.

Furthermore, if a service's security configuration needs to leverage on, or be integrated with, an existing identity and access management infrastructure then one-off integrations will need to be implemented independently on both the service and client application. Together, the complexities involved can increase development and maintenance costs in ways that counteract the benefits organisations expect of SOAs – for example, with regard to consistency, flexibility, scalability and speed of deployment.

In response to the above, new classes of security infrastructure have emerged to satisfy customer demand for purpose-built XML and web services security in both the service provider and the client environments. XML firewalls and web services gateways are dedicated devices or pieces of software that can be implemented in a socalled demilitarised zone, data centre or an application server to enforce XML and web service security behaviour based on graphical, high-level declarative policies. In some cases, they also perform hardware accelerated data transformation, intelligent routing, SLA enforcement and general purpose SOA policy operations. In all cases, they allow the enforcement of message and service-level policies with little or no programming. An increasing number of vendors offer such solutions, including Vordel [4], Layer 7 Technologies [5], IBM [6] and Cisco [7]. Such solutions are used by companies like BT to protect their service delivery platforms.

In cross-enterprise application integration scenarios, such as those common to SOI deployments [2] there is a further requirement to automate security on the client application. This is particularly important when the service imposes a policy that requires the protection of certain data elements in requests, the reconciliation of identity silos, assurance of non-repudiation or the propagation of security policy changes. While this can be done manually, the complexity of such operations and the likelihood that human errors will result in incompatibilities or weaknesses at security borders that could disrupt business operations and create substantial risks for enterprises. To mitigate them, SOA and service-oriented network security vendors such as Layer 7 Technologies offer client-side solutions (often called XML VPN) that complement their XML firewall and web services gateway products.

2.1.1 Limitations of current solutions

Although emerging XML firewall, web service gateway and VPN solutions address many of the SOA security challenges, they also present substantial limitations. Ironically, one of them is interoperability. Although XML security enforcement policies compose assertions that refer to the enforcement of widely agreed open standards, there is no agreed security enforcement policy standard for these solutions. Furthermore, those XML VPN solutions available integrate only with the XML firewalls or gateways offered by the same vendor. This increases an operator's dependency on solutions from a single vendor and limits interoperability between products.

In addition, the structure of the proprietary enforcement-policy languages offered by such products is often biased towards meta-programming of XML messaging services. Furthermore, the complexity of policies increases substantially when vendor products are integrated with external, possibly third party, policy decision points (PDP) and other value-adding services. And many of the products available are biased towards service proxy and gateway patterns: they tend to associate policies with service endpoints. This increases the complexity of administration and leads to non-intuitive enforcement policies, especially when security polices are used to control outgoing traffic such as requests to external services, where the choice of policy to apply depends on contextual factors such as location, transaction or the identity and security attributes of the requester.

Compounding these problems, commercial products are difficult to integrate with external SOA governance frameworks for policy and service assembly life-cycle management. They offer limited support for use in multiadministrative environments where policies for hosted services may be issued by different actors in the value chain. In the rest of section 2, we describe the functionality and architecture of a prototype secure messaging capability that is being developed by the authors in collaboration with some of the vendors mentioned above and attempts to address many of these limitations while preserving the benefits of current solutions.

2.2 Requirements for a policy enforcement capability

The requirements described here were elicited by the authors by studying the business and technological requirements of a large number of business cases [2], the pilot developments undertaken by research projects such as TrustCoM [8,9] and BEinGRID [10,11] and by working with SOA vendors such as IBM, Layer 7 Technologies, Microsoft and Vordel. While the requirements focus on key areas of distributed systems such as their granularity, adaptability, interoperability and scalability, they also address value-adding capabilities such as integration, decentralisation and automation.

The requirements we identified were:

- **Seamless integration** – the enforcement of security policies across distributed transactions and the seamless integration with external decision points and other value-adding services.

- **Decentralisation** – logically central policy management over a distributed policy enforcement infrastructure where the choice of the policy depends on message content and contextual information and where different aspects of a policy may be enforced by different enforcement points in the network.

- **Granularity** – separating concerns between the specification of the enforcement logic, the use of external policy decision points and other value-added services and the way that policy actions are enacted. **- Adaptability** – so that both the enforcement logic and the logic of each enforcement action can be updated at run-time

- **Interoperability** – translating internal enforcement logic into security and access requirements that clients should enforce, communicating security and access requirements to trusted clients and ensuring consistency between the internal enforcement logic and what is advertised or agreed.

- **Automation** – supporting autonomic adaptation.

- **Scalability** – coordinating a large collection of distributed enforcement points in order to secure service interactions in large-scale service oriented networks.

2.3 Anatomy of a policy enforcement capability

The authors have developed a policy-enforcement and secure-messaging capability prototype that meets the requirements described above. This prototype was initially developed in the context of the TrustCoM project [12] and is currently being extended through interactions with vendors such as Vordel and Layer 7 Technologies. The results are being validated within the scope of BEinGRID [10,13,14].

2.3.1 SOI-PEP architecture: overview

A policy enforcement point (PEP) aims to deliver adaptive, extendable policy-based message-level enforcement. The basic elements that underpin our PEP architecture for SOA policies are summarised in figure 3.

- **Enforcement middleware** – a network of service mediation and message processing nodes that intercepts each message targeted at, or originating from, a network resource or a network service endpoint. This is where service interactions are processed and service-level security policy decisions are enforced. This piece of middleware dynamically deploys a collection of message interceptors in a chain (interceptor chain) through which the message is processed prior to transmission.

- **A policy framework** consisting of interrelated configuration policies. The configuration policies constrain the type, execution conditions and order of the actions enforced on the intercepted message by the selected interceptors. The configuration policies also define which external infrastructure services can be invoked by an interceptor and the conditions of such an invocation.

- **A messaging process** model that enables selecting the appropriate configuration policies and loading and configuring at runtime the processing units that implement the enforcement actions described in the selected configuration policies. To achieve this, the processing model combines information from multiple sources including: (a) interdependent configuration policies, (b) an analysis of the structure and content of the intercepted message, (c) the current state or the history of policy enforcement and (d) contextual information that may come from the transport protocol, the content of the intercepted message or that may have to be collected by invoking external network services. An interesting feature of this processing model is that it can explore both static (pre-defined) and dynamic (emergent during processing) interdependences between configuration policies and implement them by creating multiple instances of interdependent processes that implement associated enforcement policies within the scope of executing an initial security policy.

**- A management framework** that describes the interfaces exposed by the enforcement middleware to management agents and specifies how the management agents may interact with the system.

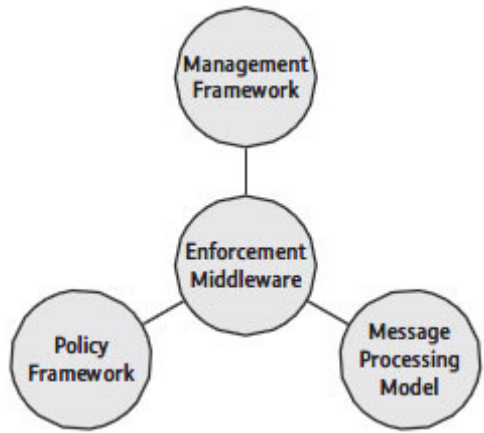

Figure 3. Enforcement framework overview

2.3.2 SOI-PEP architecture: enforcement policy framework

The enforcement middleware policy framework includes four types of policy, as shown in figure 4. The Enforcement Configuration Policies (ECPs) specify the enforcement state, the actions that are to be enacted, the conditions under which each action is executed, the parameters for each action and the sequencing of the actions. Once the relevant enforcement actions, the configuration parameters and the order in which they will form the chain have been identified, the Interceptor Reference Policy (IRP) is loaded and inspected in order to determine the references to the interceptor implementing each enforcement action. This policy maps available enforcement actions to the computational entities that execute them.

If the target executing an enforcement action requires the invocation of an external value-adding service (e.g., an external policy decision point), the IRP contains a reference identifying the external service and the Utility Service Policy (USP) used to resolve these references to the corresponding service endpoints and apply the appropriate ECP for invoking such services. If the enforcement policy dictates the use of an external value-adding service, the reference to the appropriate USP is dictated from within the relevant ECP.

While processing the message, some of the interceptors may require use of the capabilities of some external services. For example, en-/de-cryption and signature validation may require access to a key store that is external to the interceptor. Security token insertion may also require invocation of an external Security Token Service (STS)[2] that issues such a token. Similarly, security token validation may require invocation of an external STS that validates the given token. Access control enforcement may require invocation of an authorisation service that performs the access control decision.

All information regarding the alternative services available and the locations of these services are contained in the USP. This policy contains information that enables the invocation of external infrastructure services. The Capability Exposure Policy (CEP) type is used to publish additional conditions for interacting with a protected resource.

These policy types share a common meta-model that describes:

- a common endpoint reference representation for remote services or resources;

- common enforcement action types that are used in ECP and IRP; and

- common USP 'static' references that serve as rigid local identifiers of respective auxiliary infrastructure services that may need to be invoked.

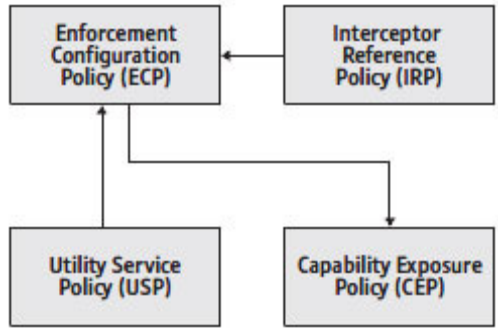

Figure 4. Separation of concerns using policies

**SOI-PEP aggregation**

To improve performance or for organisational reasons, it may be necessary to execute a policy across multiple enforcement points. Typical aggregation patterns include:

**- Cluster**: In a clustered architecture, a group of enforcement points are linked together through a master node (i.e. main enforcement point). This master node is responsible for deciding which node treats an enforcement action or a set of actions. This decision can be based on the workload of the node or the logical link(s) between different actions. In addition, the master node is accountable for keeping track of the state of the different nodes as well as currently undertaken enforcement actions.

**- In-line**: In an in-line aggregation of enforcement points, each point in the line resolves a particular part of the enforcement policy. Depending on the network topology, an in-line aggregation will bring together enforcement points deployed at the perimeter of the enterprise, the perimeter of internal administrative domains and the application deployment environment.

**Enforcement point management**

The management of the enforcement middleware is decoupled from the enforcement point itself. The overall management framework of the enforcement middleware includes enforcement point instances' life-cycle management – that is, the management of the logical association between the enforcement middleware and the current enforcement configuration, enforcement middleware configuration and management-specific notifications distribution.

The enforcement point management life cycle starts with the creation of the enforcement instance. This will be updated when relevant before finally being destroyed when it is no longer needed.

The enforcement point, which itself is virtualised as a manageable service at the control plane, exposes dedicated management interfaces to administrators and management services. These interfaces allow for policy management actions on an enforcement point such as load, activate, deactivate, destroy and roll-back to a previous successful configuration state.

2.3.3 Benefits of the solution

The core innovation underpinning this capability is a policybased, adaptive service-integration and secure-message processing layer. This layer builds on a dynamicallyconfigurable message bus is based on the best-of-breed architectures used in application firewalls, service gateways and event and service bus designs.

Another innovation is the policy-based framework that enforces security policies, performs actions that entail security policy enforcement and manages state and process isolation. Other innovations that differentiate the new solution from those currently available in the market include:

- the separation of concerns at policy specification, especially with respect to:

o policies implementing the core actions to protect a service

o policies that secure links with external policy decision points and other value-adding 'cloud' services (such as BT's 21CN services)

o policies informing service consumers of the security requirements;

- the distribution of policy actions to multiple enforcement points in a network;
- the runtime adaptation whereby security actions are loaded and configured dynamically, allowing the execution of security policies to adapt based on realtime information;
- the dynamic binding of security policies, binding the enforcement of a security policy with the policy at runtime; and
- the automatic consumer policy derivation, so that service consumers no longer have to guess what policy to apply to invoke a protected service or to develop code to implement it.

The derivation of the published CEP from the private ECP allows for an automatic derivation of security policy that service consumers have to comply with and eases the propagation of policy updates between integrated services, while maintaining secrecy of the enforcement process detail on the provider's side.

---

[a] The term Security Token Service (STS) is generally used to refer to a component that can issue, validate and/or exchange security tokens and correlate internal authentication mechanisms with standards-based security assertions about a subject.

The need for federated identity management (FIM) originates from the requirement to allow individuals to use the same personal identification to authenticate and obtain a digital identity from the networks of more than one enterprise before they can conduct transactions. Whereas FIM entails managing identities across security domains, secure federation has wider objectives and a stronger focus on infrastructure. To that extent, secure federation can be perceived as a foundation for FIM solutions that provide the interoperable service interfaces and protocols enterprises use to issue, sign, validate and exchange security tokens encapsulating claims that may include, but are not restricted to, identity and authentication-related security attributes.

A trust realm is an administered security space in which the source and target of a request can determine and agree whether particular sets of credentials provided by a source satisfy the relevant security policies of the target. If this has been established as part of the agreement, the target may defer the trust decision to a third party, thereby including that party in the trust realm.

A federation can be understood as a collection of trust realms that have established a certain degree of trust. The level of trust between them may vary[3].

Message exchanges between entities in a trust realm are typically supported by services that perform the following actions:

- the issue, validation and exchange of security-related information such as security tokens, security assertions, credentials and security attributes;
- the correlation and transformation of such securityrelated information; and
- the determination of the basis of policies that use such security information to determine an entity's entitlements for a given context and the mapping of internal authentication mechanisms to commonly agreed security attributes.

3.1 Requirements for a next-generation identity brokerage capability

In this section, we describe requirements of an identity brokerage capability for SOI that we have elicited by studying the business and technological requirements of business cases studies, research pilots, customers and SOA security vendors.

The capability should offer a framework supporting the complete life-cycle of 'constrained federations'[4] including:

- tools to manage the circle of trust that underpins an identity federation;
- tools to support 'identity bridging' between intra- and inter- enterprise identity technology, claims and authentication techniques; and
- tools to manage the full life-cycle of identities and security / access claims (provision, validation, revocation).

Trust relationships between identity brokers should be adjustable to meet the dynamics of a value chain. For example, it should be possible to correlate trust relationships between identity brokers that have supplier and/or provider links in the corresponding value chain.

The identity brokerage capability should separate concerns between the identity of the customer organisation that is using the identity broker for some collaboration, the sources of internal identities and the security administrator appointed by the customer for the selected collaboration. It should also enable each organisation to manage how security attributes and identities are assigned to their own resources in different collaboration contexts. However, the assignment of virtual identities cannot be repudiated and the provisioning of virtual identities and associated security information (such as temporary or derived cryptographic keys) should enable accountability across identity federations and throughout the value chain.

In addition, the identity brokerage capability should give management control over the selection of the schemes used for bridging internal and federated identities as well as the type and structure of information that may be conveyed within security tokens. It should also empower security managers to choose the right balance between security and privacy for their needs. It should enable them to select the degree of anonymity, the mixture of

identity-, role- or attribute-based access and the strength of non-disputable accountability of actions (e.g., end-to-end non-repudiation) they want to apply on identity brokerage in a given identity federation context.

An identity brokerage capability should be able to combine security attributes from multiple sources into standards-based security tokens (e.g., those specified in the Security Assertion Markup Language, SAML). It should be able to jointly apply recognition of authority policies, information disclosure policies and attribute transformation schemes in order to determine which security attributes have to be included in an identity token that is issued as a virtual identity for a subject within a scope of a given identity federation context.

Different security tokens may be issued or be validated depending on several contextual parameters including the issuing authority, the identity federation context within the scope of which the security token has been issued, the subject possessing the security token, policies that control information disclosure and the actions and targets for which a security token may be legitimately used by the subject possessing this token.

To facilitate integration with enterprise business processes, it should be possible to manage the identity brokerage capability remotely over a network through programmatic management interfaces and to expose it as a network-hosted (e.g., 'cloud') service that customers can integrate into their enterprise IT infrastructure.

## 3.2 Anatomy of a next-generation identity brokerage capability

An identity brokerage capability, code-named SOI-STS, has been developed that meets these requirements. The prototype was initially developed in collaboration with an innovation team from Microsoft in the context of the TrustCoM project. It has since been extended within the scope of BEinGRID, where it is being validated by customers in several business pilots and in market sectors [11].

### 3.2.1 SOI-STS architecture: external interfaces

The SOI-STS is exposed as a web service with two interfaces. Its projection on the data pane exposes an operational interface that complies with the WS-Trust standard. Its projection on the control pane is exposed through a standard web services management interface.

### 3.2.3 SOI-STS architecture: operational model

From an operational perspective, the SOI-STS architecture consists principally of the components listed in table 1. When clients request a STS to issue validate tokens, the STS will determine whether this can be done based on the information it holds in its database. If no federation context matches the request, a fault message will be returned to the requestor. Each federation context has an associated federation selector – a mechanism that maps a WS-Trust message or a management operation to an SOI-STS configuration. In a simple case, the federation selector could contain a unique identifier such as a Universally Unique IDentifier (UUID) [16] or a collection of WS-federation meta-data [18].

| STS database (repository) | A database that includes configurations of SOI-STS instances for each federation context. |
|---|---|
| Federation module | A module associated with each (class of) federation context, consisting of a federation selector and delegation constraints. |
| Federation | A capability of the SOI-STS that maintains local state describing |
| partner provider | the trust relationships of the STS in each federation context known to the STS. It also allows the other STS components and processes to retrieve information about a circle-of-trust that is identified by a unique 'federation identifier'. This information will typically include security information to identify the STS of each trusted federation partner and be able to validate identities issued by this STS. It may also include information indicating the level of trust in this federation partner and potential constraints on how to process information provided by or disclosed to this federation partner. |
| Claims provider | A capability of the SOI-STS that provides a set of claims for a given 'internal' identity. It may also maintain a catalogue of internal identity providers and information about their association with the STS. This capability will typically be used during a token issuance process. It may also apply potential constraints about a federation context and/or an 'internal' identity. |
| Claims validity provider | A capability of the SOI-STS that maintains associations between federation contexts, security token types and policies that determine the validity of security claims. |
| Claims transformation provider | A supplementary service that applies a rules-based transformation between taxonomies of 'internal' and 'external' security attributes. |
| Authentication scheme selector | An auxiliary service that selects the mechanism used to authenticate an entity requesting the issuance of a token and generate the associated 'proof-of-possession' information. |
| Service access provider | A possible extension to the claims validation provider service that allows integration with, or incorporation of, the functionality of an authorisation service. |
| Obligation policy provider | An auxiliary service that can provide 'obligation policies' that offer, for example, associated policy enforcement points and instructions about further actions to be performed in order to complete a token issuance. |
| STS business logic | This defines a process that uses the internal component services mentioned above and is executed in order to issue, validate or exchange a token in response to a well-formed request. |

Table 1. SOI-STS architecture main components

---

[a] In [3], we analysed the basic architectural concepts that underpin trust realms and their federation.

[b] Constraint federations are federations of trust realms where trust relationships between the federating parties may vary depending on a set of constraints about recognition of authority.

After selecting the matching federation
configuration, an STS will instantiate the STS business logic provider and load it with the applicable process description. It will also instantiate the internal capabilities of the STS such as the federation partner provider, the claims provider and the claims validity provider and bind them to the STS business logic process. Each of these capabilities of the STS may have a federation-context-specific configuration, which will be loaded upon instantiation.

3.2.4 SOI-STS architecture: management model

To manage a set of dynamically instantiated services as pluggable modules, the SOI-STS management interface is split into two parts: a set of 'core' management methods and a single 'manage' action that dispatches management requests to dynamically selected modules. The signature of the 'manage' method depends on the modules integrated in a given instance of the SOI-STS. The flexibility of XML and SOA web services technology accommodates this form of dynamic composition.

Referring to figure 5, the core management methods include operations for creating new federation configurations from given specifications, for temporarily disabling or enabling them and for inspecting their values and meta-data. A proxy function forwards provider-specific management requests to the respective provider management module.

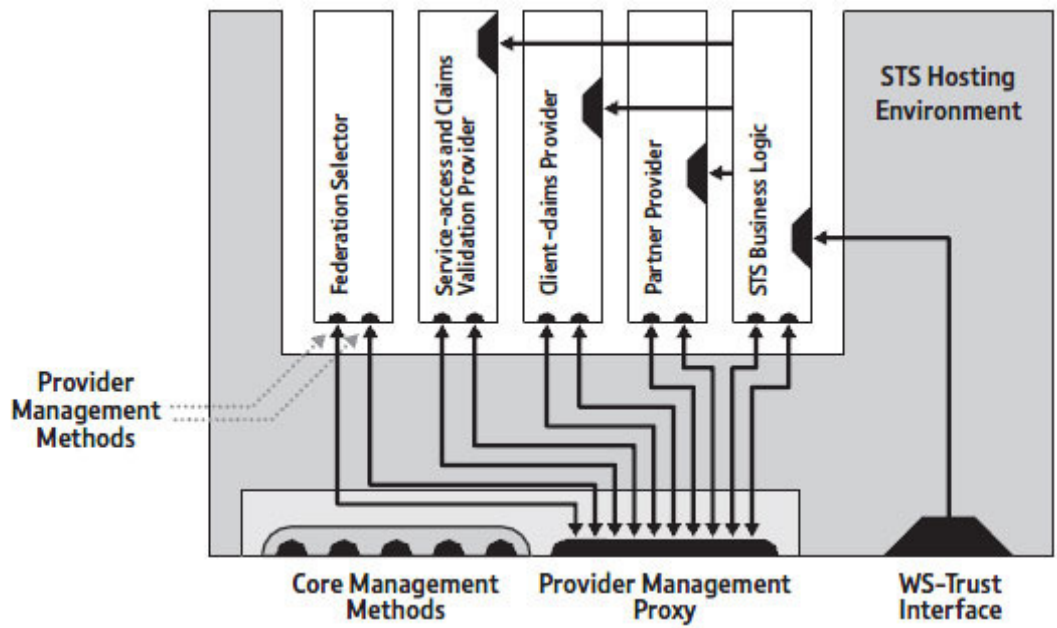

Figure 5. Management model

3.2.5 Benefits of the solution

Identity brokerage capability is best understood as an environment that allows the assembly and runtime provisioning of identity broker instances that adapt their trust models and processes based on the context within which brokerage is required.

The design of the identity brokerage capability balances agility and extensibility with ease of integration and compliance to standards:

- Ease-of-integration: The core operational and management interfaces are static and implement widely accepted standards;
- Agility: The solution enables the definition of contextspecific identity brokerage processes and policies about how information is managed in a context. Once an identity brokerage request is received, these processes and policies are bound together into a context-specific identity broker execution thread that is provisioned at runtime.
- Extensibility: The solution enables the introduction of new modules, such as the integration of 21CN and other 'cloud' services into the identity broker definition environment.

This capability is well suited for Identity-as-a-Service (IaaS) propositions. It enables the context-based virtualisation of identity brokerage services and supports their remote management and federation.
trust network that can reflect the service-consumer relationships for a particular value network and a particular context. Brokers can share the same federation context identifier (i.e., a shared state reference) and associate it with their internal view of the circle-of-trust that reflects their own trust relationships (i.e., local state). The latter may include assertions recognising the authority of those identity brokers they trust in this federation context. Directed binary trust relationships can be defined between an identity broker and each of the trusted identity brokers with which it is associated in a federation context. This model can be further extended by including a representation of trust metrics such as those proposed in [19] (see also [20]). Depending on the distribution of recognition-of-authority statements, trust relationships in such trust networks may be adjusted to reflect the value chain of the corresponding business-to-business collaboration. For example if IB1 is a prime contractor recognising the authority of subcontractors IB2 and IB3 in federation context F1 and each of IB2 and IB3 recognise only the authority of prime contract IB1 in F1 then IB1 will be able to process the validity of tokens issued by any of IB1, IB2, IB3, while either of IB2 and IB3 will be able to process the validity of tokens issued by IB1 and itself only. This is summarised in figure 6.

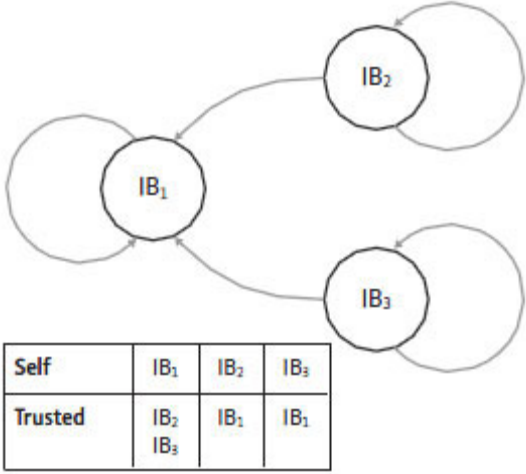

Figure 6. Example of an asymmetric circle of trust

The identity token issuance and validation are contextualised: different virtual identities and entitlements can be issued for the same internal identity, depending on the context of the issuance request and different validation results can be obtained for the same token depending on the context of the validation request.

| Role-based | RBAC is an authorisation mechanism that associates a set of |
| access control (RBAC) | access privileges with a particular role, often corresponding to a job function (e.g., finance director, student). It simplifies security management by providing a role hierarchy structure whereby one role can inherit rights from another and thus avoid repeating the specification of permissions. More recent development in RBAC has seen the introduction of constraints to restrict the assignment of users or permissions to roles or the activation of roles in sessions. |
| Attribute-based access control (ABAC) | ABAC provides a mechanism for defining permissions based on just about any security-relevant characteristics, known as attributes. A subject's (e.g., a user's or an application's) access profile is defined through a combination of the following attribute types:<br><br>- Subject attributes defining the identity and characteristics of a subject<br>- Resource attributes associated with a web service, system function or data<br>- Action attributes associated with a resource's possible actions, they can restrict what action can be invoked on the desired resource<br>- Environment attributes describing the operational, technical or situational environment or context in which the information access occurs.<br><br>ABAC policy rules can be custom-defined with consideration for semantic context and are significantly more flexible than RBAC for fine-grained alterations or adjustments to a subject's access profile. |
| Policy-based access control (PBAC) | PBAC introduces the notion of a policy authority, which serves as the access decision point for the environment in question. PBAC leverages the granular policy rule functions inherent to ABAC. |

Table 2. Main authorisation mechanism styles

The identity brokerage capability architecture balances extensibility and compliance to standards: the core operational and management interfaces implement widely accepted standards, while extensibility is facilitated by enabling the introduction of new modules implementing mission-specific identity and security management models.

Distributed access control and authorisation services allow groups of service-level access policies to be enforced in a multi-administrative environment while ensuring regulatory compliance, accountability and auditability. Until recently most of the research into access control for networks, services, applications and databases was focused on a single administrative domain and the hierarchical domain structures typical of traditional monolithic enterprises. A brief summary of the outputs of this research is presented in table 2.

The dynamic nature and level of distribution of the business models that are created from an SOI [2] often mean that one cannot rely on a set of known users (or fixed organisational structures) with access to only a set of known systems. Furthermore, access control policies need to take account of the operational context such as transactions (for example, as identified in specific WS-A message IDs) and threat levels. The complexity and dynamic and multi-administrative nature of an SOI necessitate a rethink of traditional models for access control and the development of new models that cater for these characteristics of the infrastructure while combining the best features from RBAC, ABAC and PBAC.

4.1 Requirements for an access management capability for SOI

As with policy enforcement and identity brokerage, requirements were identified by studying the business and technological requirements of a large number of business case studies, research pilots created by projects such as TrustCoM [8,9] and BEinGRID [11], by working together with customers [21] and through interactions with SOA vendors such as Axiomatics, IBM, Microsoft and SAP.

The access management capability should enable the necessary decision making for enforcing groups of service-level access policies in a multi-administrative environment while ensuring regulatory compliance, accountability and security auditing. It should be able to recognise multiple administrative authorities, admit and combine policies issued by these authorities, establish their authenticity and integrity and ensure accountability of policy authoring, including the nonrepudiation of policy issuance. The validity of the access policies issued may be time-limited and must be historically attested.

The access management capability should also cater for policies addressing complementary concerns (operational and management) in a multi-administrative environment. It should support policies about:

- Subjects accessing resources in a context, where policies will be issued (and signed) by administrators authorised to manage resources.
- Who can delegate which access rights about which resources and in what context.
- Obligations that instruct associated policy enforcement points. An advantage of using a policy-based enforcement point is that obligations may include references to CEP assertions, thereby providing semantically-clear instructions to the enforcement point.
- Administrative delegation, which defines who can author access policies and constraining the applicability of policies authored by administrative authorities. Constraints may take the form of rules that apply on a subset of the available attribute types and policy evaluation algorithms.

In all cases, there may not be any prior knowledge of the specific characteristics of subjects, actions, resources and so on. Hence, there are no inherent implicit assumptions about pre-existing organisational structures or resource or attribute assignment. This comes in contrast to access control lists and traditional role-based access control frameworks. During access policy evaluation, access decisions may need to consider environmental attributes and other contextual information in addition to subject, resource and action attributes. Policy administration and decision making may also be contextualised. Different administration and/or command structures may manage independent life-cycle models and policy groups associated with different contexts. Access policies may also need to be executed within the scope of a particular context that influences the way in which their evaluation algorithms are being applied. In some cases, it may also be necessary to ensure segregation of policy execution – that is, that there is no interference between the policies being executed in different contexts.

The policy decision point (PDP) at the core of the access management capability may be exposed as a hosted service, be deployed as a component of a policy decision making capability with a larger scope (such as a federated identity and access management capability) or be an integral part of the policy enforcement (PEP) function. It should also be possible to deploy the overall access management capability as a managed service, if needed.

4.2 Anatomy of an access management capability

A prototype identity brokerage capability called SOI-AuthZPDP has been developed that meets the requirements described above. It was initially developed in collaboration with a research team from the Swedish Institute of Computer Science (SICS) in the context of the TrustCoM project and has been extended within the scope of BEinGRID, where it has been validated in business pilots involving market sectors from media and entertainment to engineering and e-health. Ongoing improvements are being developed in collaboration with a spin-off from the SICS's Security, Policy and Trust (SPOT) laboratory and Axiomatics [22], which aims to commercialise the prototype.

4.2.1 SOI-AuthZ-PDP architecture: external interfaces SOI-AuthZ-PDP exposes three interfaces:

- An administration interface, called the Policy Administration Point (PAP), which is typically exposed as a web service complying with service management standards and accepting policies in standardised accesscontrol languages such as the current XACML standard [23,24]. The management interactions are detailed later in this section.
- An attribute retrieval interface that joins together adaptors to external attribute authorities. - An operational interface. Depending on the form of deployment this can be a web service implementing standard access control queries such as the XACML request profiles that have standardised bindings over the Simple Object Access Protocol (SOAP) and a SAML profile [25]. The operational interactions are detailed later in this section.

4.2.2 SOI-AuthZ-PDP architecture: data structures

The core elements of the information model include the policy issuer, the policy target, the policies and the policy decision request and response.

The policy issuer is an identifiable entity that has the authority to provision access policies (including entitlements). The policy issuer may have certain entitlements about the kind of policy targets and policies that it can author and all policies issued should be signed by the corresponding policy issuer. The policy target is the collection of variables on which a policy would apply. In the case of access management policies, these may include attributes identifying some subjects, resources and actions on resources. Environmental variables such as time, transaction context and so on may also be taken into consideration.

Policies are collections of rules and constraints that apply on one or more policy targets. In the case of access management, policies will typically be about what kind of actions some identifiable subjects can make on some identifiable resources within a scope that is characterised by some environmental variables. Policies are combined in policy groups by means of policy combination algorithms that serve to resolve conflicts by prioritising and overriding among policies that apply on overlapping policy targets. The policies that underpin the prototype access management capability fall in three categories: root policies, delegated policies and administrative policies. These are used together in a process of validating constrained delegation of administrative authority in multiadministrative environments.

Constrained delegation validation is a process that involves looking for root policies which authorise the delegated policies in accordance to the constraints defined in the administrative policies.

Root policies are signed policies or policy sets. They are stored in a different compartment of the policy store than the delegated policies. When SOI-AuthZ-PDP loads a root policy, it will not generate a policy issuer, which must be among a collection of pre-configured trusted authorities that are established without delegation validation. The root policies are used to verify the authority of signed delegated policies.
Delegated policies are signed digitally by the administrative authority that issues them – that is, by the corresponding policy issuer. They are stored in a special compartment of the policy store. When SOI-AuthZ-PDP loads a delegated policy, it will use the digital signature to establish the policy's authenticity and generate a policy issuer description and associated validity constraints. The policy issuer will result in the PDP performing constrained delegation validation on the policy before it is used. Administrative policies define the constraints that inform the administrative delegation.

Normative policy administration should happen through signed policies. The root policies define the authority of normative policy administration.

In our research prototype, the SOI-AuthZ-PDP responds to an XACML access request by locating a group of XACML policies that apply to that request. Delegated policies are validated according to the XACML 3.0 administrative delegation model before they are used. Administrative policies define the constraints that inform the administrative delegation. The validation involves looking for root policies which authorise the delegated policies in accordance to the constraints defined in the administrative policies. The root policies and the delegated policies are grouped together into an XACML policy set by applying policy combination algorithms that resolve conflicts by overriding behaviour [26].

'Overriding' means that policies in a policy group are classified and prioritised and they are all jointly applied on a given target. Decisions of higher priority policies override the lower priority policies whenever there is an overlap between policies in a group within the scope of the given target. Overriding is localised on the overlapping area. All policies in the policy group apply equally outside of the overlapping area.

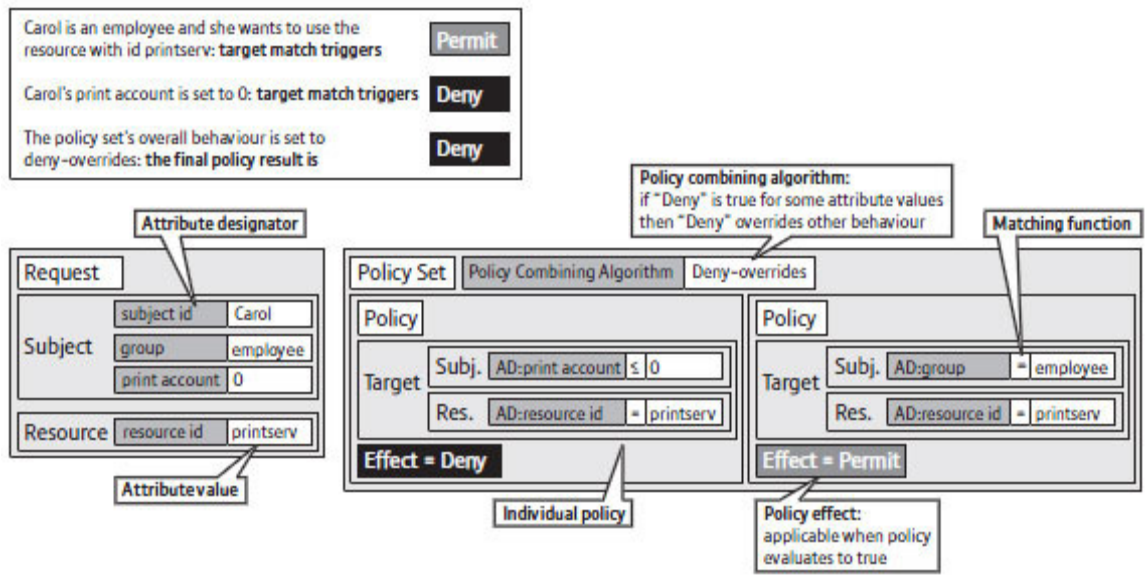

Figure 7. An example of access request and policy combination

An example of a policy request and an applicable policy set is shown in figure 7. Additional administrative delegation constraints may be used to scope the evaluation of the policy-set.

4.2.3 SOI-AuthZ-PDP architecture: operational model

From an operational perspective, the SOI-AuthZ-PDP architecture is as shown in figure 8. A requester (e.g., the end user in the figure) uses an application that contains or is deployed in a Policy Enforcement Point (PEP). The PEP will intercept any attempted use of the application and generate an XACML request that describes the attempted access in terms of attributes of the subject, resource, action and environment. The request is sent to the PDP. The PDP will process the request and send back an XACML response with a permit, not applicable or deny decision, or a decision indicating an error condition and (optionally) obligations. The PEP will enforce the decision and let the subject access the resource or block the access depending on the decision. The PEP will also enforce any obligations contained in the response. The query pre-processor indexes the XACML query into a form which is efficient to process and generates individual queries in case the incoming request concerns multiple resources. The query pre-processor may also optimise multiple resource requests by invoking partial evaluation of XACML policies.

The XACML evaluator evaluates the query using the XACML function modules. The XACML evaluator may retrieve additional external attributes which were not present in the incoming XACML request.

The loaded policies are indexed in an efficient form in live memory, where the query pre-processor and the XACML evaluator will retrieve the policy form for evaluation. Attributes could be stored locally or be obtained during policy evaluation from an external repository (LDAP directories for instance).

When the PDP receives an XACML request, the query pre-processor will parse, validate and index the request for processing. The query pre-processor will get the current policy of the PDP and optionally optimise the request by calculating an optimised policy. The policy is then evaluated and a decision sent back to the PEP. During evaluation, PDPs can retrieve additional attributes from external sources.

4.2.4 SOI-PDP architecture: management model

From a management perspective, the SOI-AuthZ-PDP architecture consists of the following main components:

- A service acting as the Policy Administration Point (PAP). This is the entry point for policy administration and service management. A policy administrator uses the PAP (by a GUI client or programmatically) to administer the policies in the policy store. Access to the policy store is done through a PAP service which enforces invariants and access control on the policy repository. The PAP service will also perform access control on the policy store and will make the required changes in the store. The PAP service will consider administration of the root policies to be a sensitive management operation which is protected by stricter access control than administration of delegated policies.
- A PAP client that offers graphical interfaces for administrators to view the policy store and its contents and perform common operations such as adding, removing, changing, signing and activating policies and so on. The client interacts with the PAP service through a web services interface and web service management protocols.
- Attributes and policies stored locally in attribute and policy stores or in a distributed manner (using LDAP directories, for instance).
- A policy loader that loads policies from the policy store.
- A policy validator that can be used by the policy loader to validate the policies syntactically, verify the digital signature on the policies and, in case of delegated policies, generate the XACML policy issuer from the signature and amend any applicable administrative delegation constraints.

At the time of writing, we were working with vendors to implement a range of extensions to the SOI-AuthZ-PDP architecture (see [23]).

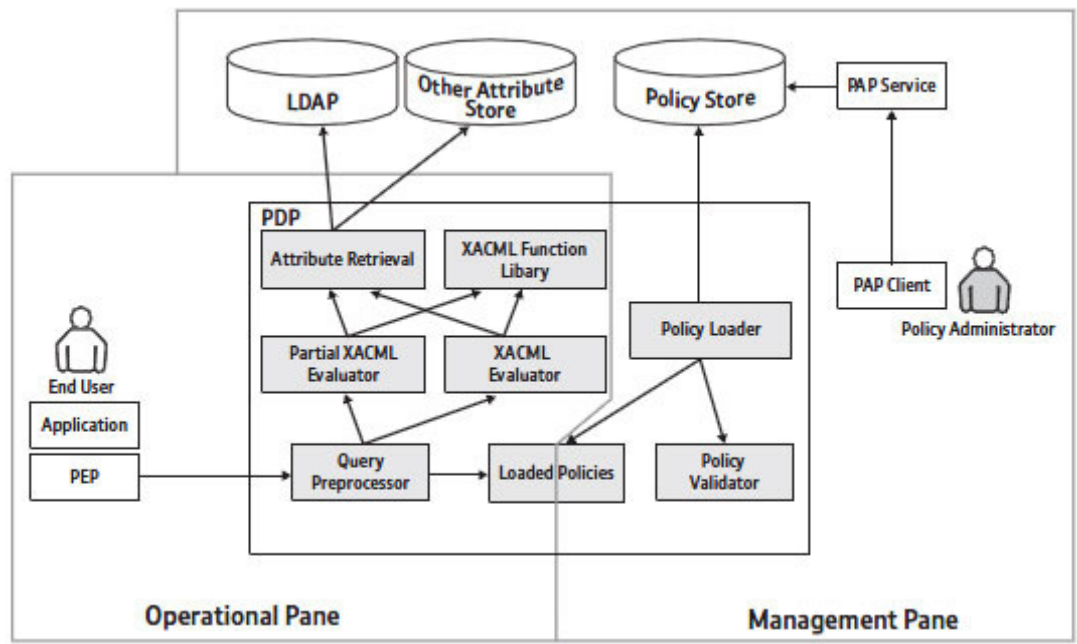

Figure 8. PDP – internal functional components and external interactions

4.2.5 Benefits of the solution

An access management capability has been developed that combines logically-centralised policy management with the agility of decentralised administration of access policies and the option of distributing policy enforcement. Compared to other access management capabilities, its benefits become more prominent in scenarios relating to multi-administrative environments (e.g., business, government or defence coalitions), to shared IT and communications infrastructures and to multi-tenancy hosting.

One of the scenarios to which the solution can be applied includes a defence coalition in which local administrators must act quickly using their local knowledge to define policies within the scope of constraints agreed by the coalition's command and control. A second class involves the sharing of resources across multiple organisations or organisational units that wish to keep some control over the shared resources. In the third class, an IT or communications provider hosts services or information owned by its customers or offers core capabilities that enable its customers to assemble and offer their own customised services.

The innovations that differentiate the solution from other access management capabilities include:

- **Delegation of administrative authority**: policy authoring and management is controlled by constraint-delegation policies that put constraints on the access management policies that administrators can author and allow the runtime creation of dynamic chains of delegation of administrative authority without assuming prior knowledge of an organisation's structure. These constraints restrict the validity and scope of access management policies during run-time policy decision making.
- **Authenticity, integrity and accountability of access policy** authoring: policy authoring rights are granted to issuers whose accountability is enforced by use of digital signatures, policy issuer identities and evidence gathering.
- **Context-based capability virtualisation**: policy stores and policy-execution processes can be segregated based on contextual information.

Additional benefits result from the use of a flexible, standards-based policy language and the versatility of deployment. A prototype of this capability implements XACML thus improving reuse, rapid customisation and interoperability. Unlike most of the currently available implementations of XACML, it leverages the concepts of constraint delegation and policy issuance that are being introduced in the current draft v3.0 of the OASIS standard. Furthermore, the core functions of the capability can be exposed as a network-hosted (a.k.a. 'cloud') service or deployed as a component linked into a service gateway, an enforcement layer, a service container or the application itself.

The high-level functionality of the novel security solutions developed by BT in collaboration with academic researchers and product vendors is summarised in table 3, while relationships to different aspects of SOI and policy management are shown in figure 9.
A wide spectrum of complementary concerns is covered by the capabilities that have been developed, including:

- ensuring the secure exposure and availability of services;
- protecting the confidentiality and integrity of the information and data exchanged between these services;
- managing service-level access in multi-administrative environments;
- brokering and federating identities;

- managing trust in B2B collaborations;
- governing the distribution security policies and the composition; and
- managing security capabilities that are deployed within the enterprise or are hosted by third parties.

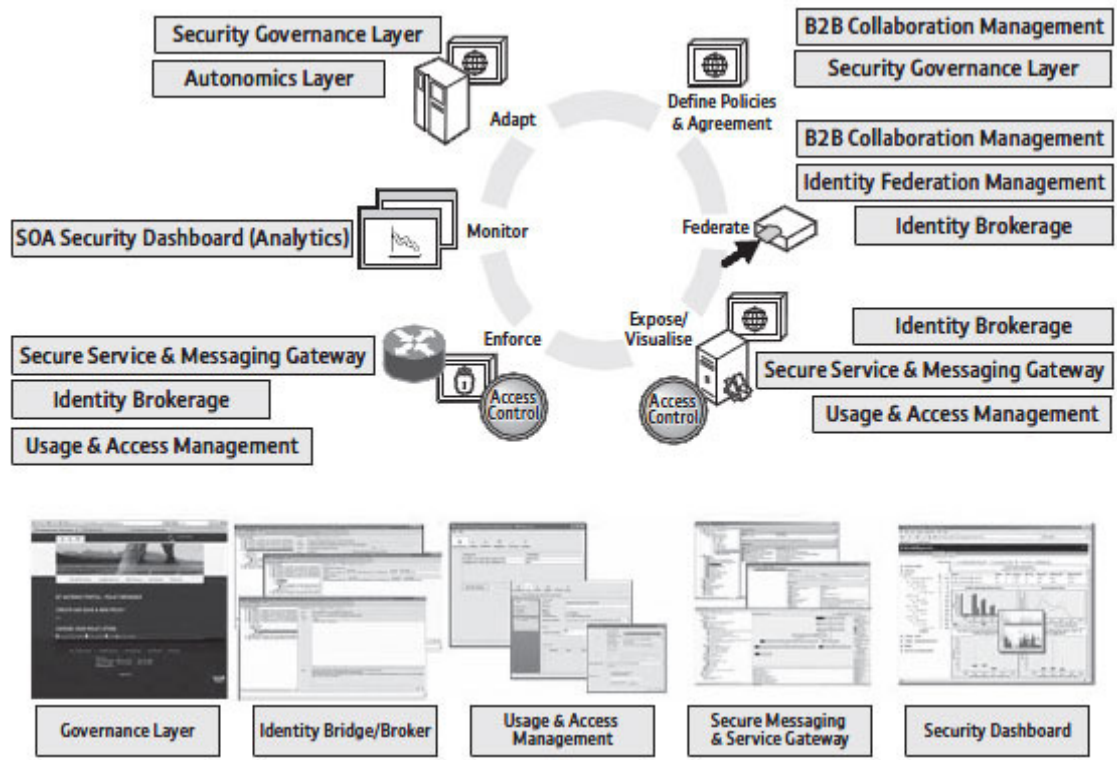

Figure 9. Overview of SOA security common capabilities

Each of these capabilities can be deployed as a web service with its own service management and policy administration framework (control pane) and operational interfaces (data pane). Standards-based programmatic interfaces facilitate integration with Enterprise Service Bus (ESB) and other third-party SOA governance tools.

The capabilities can also be composed into a secure service gateway for the SOI (code-named SOI-SSG), which secures the exposure and end-to-end integration of business applications within the enterprise, between the enterprise and its customers and among business partners. SOI-SMG can offer the security subsystem of a service delivery platform or a service gateway ensuring that corporate applications and platforms can securely access specific enterprise functions over public networks. It can also be used for securely exposing value-adding services such as BT's 21CN common capabilities [27] or some of the reusable services at [10] to a network. When used in conjunction with SOAbased service integration platforms, SOI-SSG enables the seamless integration of value-add services into the corporate ESB.
There are two main integration points when composing such capabilities:

- **A SOA security governance layer** that manages the service exposure life-cycle and coordinates the PAPs of the security capabilities integrated in the SOI-SMG.
- **A network of PEPs** that integrates the operational interfaces of the SOI security capabilities and the protected business services.

| Security capability | Functionality |
|---|---|
| B2B collaboration management | A collection of managed services that support the full life-cycle of defining, establishing, amending and dissolving collaborations bringing together a circle of trust (federation) of business partners. For a description of these services, please refer to [12] and [17]. |
| Identity brokerage SOI-STS | A policy-based and context-aware identity broker that allows the representation of federation contexts (circles of trusts) and can issue, validate and exchange virtual identities (security tokens) while (a) implementing different virtual identity schemes, credential mappings and authentication mechanisms; and (b) recognising different external identity authorities depending on this context. |
| Identity federation management SOI-FMS | Facilitates the management of full life-cycle of circles of trust, by coordinating a distributed process that establishes trust between the participating partners. Allows the creation of trust relationships between STS instances that reflect the dynamics of supply chains. |
| Usage/access management SOI-AuthZ-PDP | An authorisation service that automates usage and access management decision-making based on access management policies that can be authored by multiple administrators, while facilitating the composition of policies from different administrative authorities, with policy analysis to prove regulatory compliance and accountability and security audit of administrative actions. |
| Secure service & messaging gateway SOI-SMG | A fusion of (a) an application service firewall / gateway that protects interactions to XML applications and Web Services, (b) a proxy that intercepts, inspects, authorises and transforms content on outgoing requests to external services, (c) a message bus that enforces content- and context- aware message processing policies and (d) a light-weight core of a service bus that integrates the interfaces exposed by all other SOI security capabilities in the data pane. |
| Analytics (SOA security dashboard) | A collection of services that correlates and analyses notifications representing events reported by the other security capabilities. It may perform complex event processing in order to identify and classify a security or reliability event based on the events reported and may also perform risk analysis. |
| Autonomics layer | A collection of services to reconfigure the security services based on declarative adaptation policies and in response to security or QoS events in order to optimise performance, to respond to threats and to assure compliance with agreements and enterprise policies. |
| Security governance layer (SOA security governance gateway) | A governance layer managing (a) the life-cycle of a secure exposure of business services, (b) the composition of such services with a collection of SOI security capabilities that implement nonfunctional requirements and (c) the life-cycle of policies associated with each SOI security capability in order to implement non-functional requirements associating with the exposure of a business service. |

Table 3. Summary of functionality offered by different security capabilities

Figure 10 shows how SOI security capabilities can be composed into SOI-SMG during service operation. The secure Service and Messaging Gateway (SMG) intercepts messages addressed to a set of resources and enforces a security policy and integrates additional security capabilities (such as identity brokerage, authorisation services and other managed security services) to secure the resources' communications.

Policies can be contextualised: depending on the context of the request, the SMG will load a certain set of these policies. Context may depend on meta-data such as the location of the requestor, the region of the endpoint being invoked, references to business transaction types in the message, state of alert, etc.

In figure 11, the SOI-SMG is integrated with identity brokerage and authorisation decision points. In accordance with the selected policy, the SMG will request an XML token on behalf of the requestor at the client side, passing on the appropriate context reference to an identity brokerage capability. The identity broker will inspect the token issuance request and any associated context references and will select the appropriate identity providers, attribute transformation schemes, identity federation parameters and security token schemes. The selection translates internal identity, such as an X509 binary certificate or a Kerberos ticket, to a set of commonly agreed security assertions that are aggregated into a security token (for example, one following the SAML standard) that acts as a temporary virtual identity. The identity broker should also generate a proof-of-possession key – cryptographic material used by the SOI-SMG to ensure that the integrity and authenticity of the requests to the service can be proven. This token will be embedded into the outgoing request and certain message elements will be signed with the proof-of-possession key provided by the identity broker in order to establish the authenticity of the data contained in these elements and bind them to the provisioned identity. Depending on the scenario, the same, a derived or a different cryptographic material may be provisioned for encrypting message elements in order to ensure end-to-end confidentiality of data targeting specific recipients. Although providing cryptographic material to ensure data confidentiality is not part of the normative operation of an identity broker, the architecture of the STS described in section 3 facilitates such enhancements if they are needed.

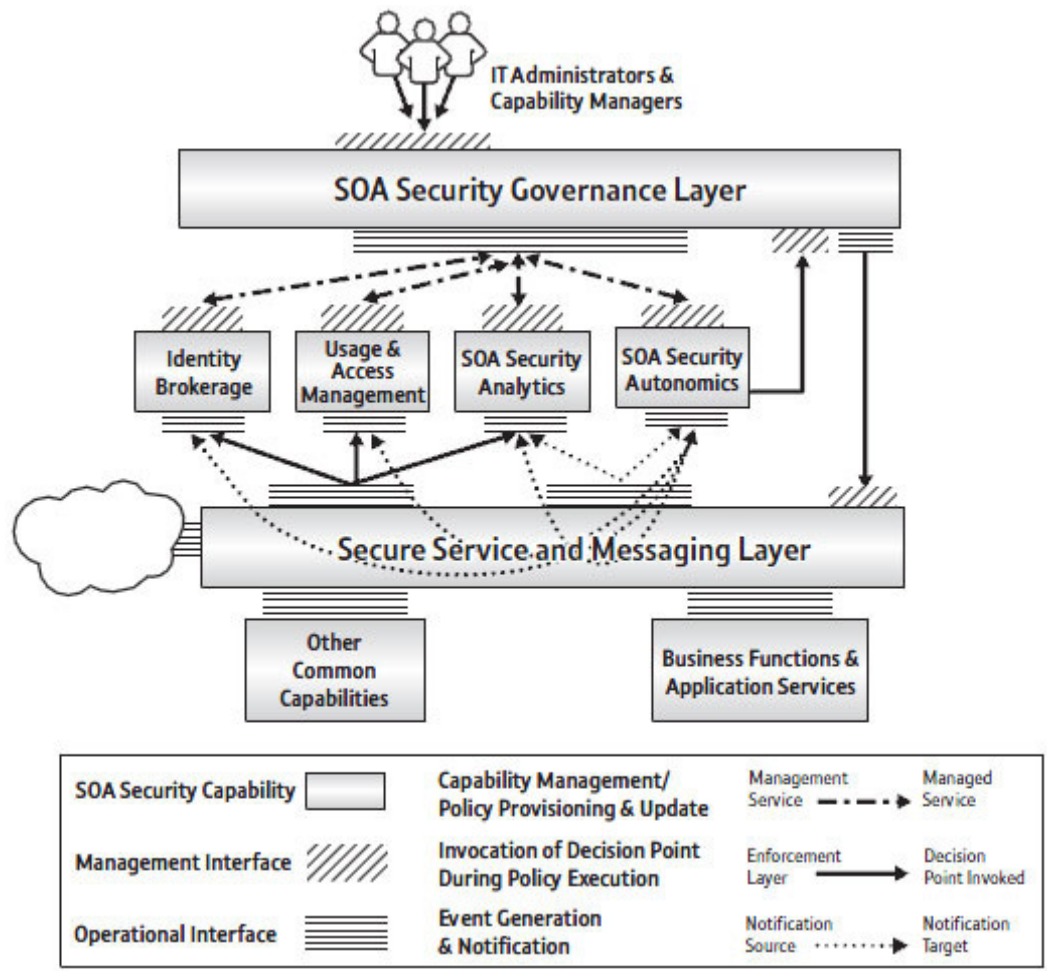

Figure 10. SOI-SSG integration

On the service-side, the SMG of the targeted partner performs any message processing and protection actions required by the policy and will also extract the XML token from the incoming message and send it for validation at its own local identity broker. The latter will apply the appropriate token and claims validation procedure and will inform the SMG whether the token is valid and if so provide the list of associated claims.

The SMG can then use the claims for authorisation queries to determine whether the requestor is allowed to use the action specified by the incoming request on the targeted resource. Authorisation decisions may also depend on contextual references and other information collection from the operational environment. For every run-time action, the SMG can generate monitoring and audit events. In addition to security auditing, such events can be used by the autonomics layer of the SMG in order to optimise its configuration or respond to changes of the context of the interactions. The autonomics layer processes the events, correlates them and determines whether to produce other events (e.g., an alert or a reconfiguration notification to another SMG node) or to trigger reconfiguration actions by invoking a management process via the governance layer.

An example of an adaptation event is the case where a targeted partner repeatedly receives requests with valid XML tokens, therefore ensuring the request does come from an authenticated and recognised requestor, but with invalid claims and attributes forcing the authorisation service into denying access to the desired resource. As a result the SSG will fire off an event to the adaptation engine. After a set threshold, the adaptation service can ask the partner that issues the XML authentication token to reconfigure its infrastructure to ensure that clients either have the appropriate claims in the future or be prevented from making calls altogether.

In this paper, we have provided an overview of concepts, models and technologies that can be used to secure operations in service-oriented enterprises. Examples from an SOI security framework developed by BT were used to illustrate how the concepts and technologies can be combined to achieve security in IT-driven business environments and illustrate the provision and control of security services in a service-oriented world. The analysis and results presented stem from ongoing research. A number of solutions have already been patented that support the realisation of SOA through a collection of context-sensitive, policy-based and service-oriented security capabilities, and a complementary collection of design patterns that support the composition of these capabilities into secure SOA blue-prints that are fine-tuned to secure business operations in different contexts.

One area for further research is the improvement of policy-based management of the 'circles-of-trust' that underpin trust in virtual organisations or other forms of business-to-business collaboration. This capability, together with identity brokerage, could be offered as a networkhosted service, extending the Identity as a Service (IaaS) provisioning model [28]. (See also [29] for an example of a recent IaaS solution.)

Another area that needs further research is the development of the secure SOA governance layer to improve the support it provides for assembling SOA security capabilities into secure SOA profiles within different business contexts and for coordinating service and policy management throughout the SOA life-cycle.

We also plan to explore opportunities for innovations developed to date to be built into the SOA vendor products that BT and others use to ensure the security and agility both of their own operations and of those of their customers. Finally, further work will be done to validate the solutions developed in the different business application contexts typical of virtual engineering, retail, e-health and defence organisations and of service providers operating virtualised ICT infrastructures.

The authors acknowledge Srijith Krishnan Nair and Afnan Ullah Khan at BT Research for their significant contribution to the development and improvement of the work presented in this paper.

The contributions made by the partners in the European collaborative research projects TrustCOM and BEinGRID are also acknowledged, along with those made by the researchers involved in BT's service-oriented infrastructures research programme.

In particular, the authors would like to acknowledge Dr Christian Geuer-Pollmann and Dr Joris Claessens of Microsoft European Innovation Centre for their contribution in designing and developing an earlier prototype of an identity broker in the TrustCoM collaborative research project, and Erik Rissanen, Dr Babak Sadighi and their teams at SICS and Axiomatics for their contributions in the area of entitlement and access management.

We also acknowledge Mark O'Neil, Richard Mooney and their team at Vordel and Phil Watson and Francois Lascelles and their team at Layer 7 Technologies for their contribution in the area XML gateways and policy enforcement. Early stages of this research were benefited by discussions with academic partners including Dr Emil Lupu at Imperial College, Professor David Chadwick at the University of Kent and Dr Panos Periorellis at the University of Newcastle. Last but not least, the authors acknowledge Dr Ivan Djordjevic (now at CA), Dr Leonid Titkov (now at HP) and Andreas Maierhofer for the contributions they made to the work presented in this paper while at BT.

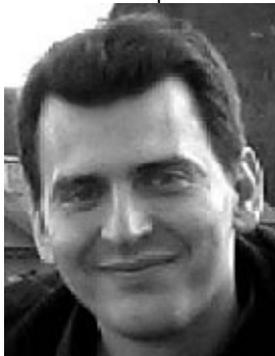Theo Dimitrakos graduated from the University of Crete in 1993 and gained a PhD from Imperial College, London, in 1998. With fifteen years experience in a wide range of topics relating to information security and software and systems engineering, he now leads the SOA Security Research Group in BT's centre for information and security systems research. Theo also has a strong academic background in the areas of security risk analysis, formal modelling and applications of semantics and logic in computer science. He was the scientific coordinator of the European Union's BEinGRID and TrustCoM research projects and contributed to a UK Department of Trade & Industry Foresight project on cyber trust and crime prevention. The author of more than fifty scientific papers and (co-)editor of five books and two special editions of technical journals relevant to his interests, Theo serves as vice-chair of an International Federation for Information Processing (IFIP) working group on trust management and a member of the IFIP special interest group on enterprise interoperability.

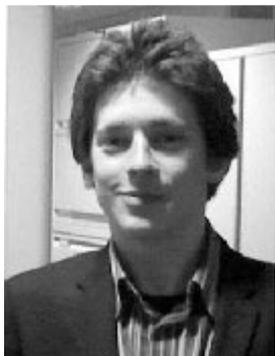David Brossard received his MEng from the Institut National des Sciences Appliquï¿½es in Lyon, France. He is currently a senior researcher in the security architectures group of BT's Centre for Information Systems & Security Research, where his interests centre on SOA security and governance. David is a Suncertified enterprise architect, an affiliate of the Institute of Information Security

Professionals, a member of the IEEE and is working towards achieving CISSP accreditation. He has been actively involved in European Union research projects including TrustCoM and BEinGRID, where he leads the security theme. Prior to joining BT, David worked as architect/designer at Portugalmail, a Portuguese ISP, working on the company's blogging platform. He also worked for leading defence company, Thales, in various programming roles.

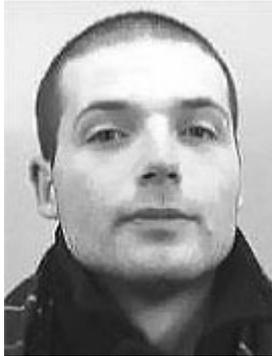Pierre de Leusse graduated from the University of Teesside in 2004 and gained an MSc from the university the following year. Currently, he is a PhD student in Newcastle University's distributed systems group researching architectures for SOA governance that allow the secured contextualisation and adaptation of web services. Previously, while working at the University of Teesside as a researcher/ consultant, he developed an ontology-based framework for the automated discovery of web services. He also managed a number of SOAbased knowledge transfer projects involving the university and nearby companies. The projects were designed mainly to help companies improve the scalability and adaptability of their software solutions.